# Realization of Ohmic-contact and velocity saturation in organic field-effect transistors by crystallized monolayer

Boyu Peng[1], Ho Yuen Lau[1], Ming Chen[1], and Paddy K. L. Chan[1*]

1. Department of Mechanical Engineering, The University of Hong Kong, Pokfulam Road, Hong Kong

Organic semiconductor (OSC) materials are the essential component in various emerging applications, including organic field-effect transistors (OFETs), organic light-emitting diodes (OLEDs), and organic photovoltaics (OPVs). In all of these devices, the efficient charge transport across the metal-semiconductor (M/S) interfaces determines the device performance to a great extent, for example the on-state current and the transit frequency of transistors, the external quantum efficiency (EQE) of OLEDs, and the power conversion efficiency of solar cells. Generally, when the metal forms a contact with the semiconductor, a Schottky barrier is usually induced due to the misalignment between the metal work function and the transport energy levels in the semiconductor, and this Schottky barrier blocks the current flow across the metal-semiconductor interface to a certain extent.[1] This issue gets particularly severe at metal-OSC (M/OSC) interfaces due to the abrupt difference in carrier density, the large density of interfacial traps, the Fermi-level pinning effect, etc.[2] The Schottky M/OSC contacts induce potential drops at the contacts which restrict the high-power operation, or result in non-linear charge injection as well as device-to-device variations.[3-5] For OFETs, the Schottky contacts would cause large (>1 kΩ·cm) and voltage-dependent contact resistance ($R_c$), which become the major bottlenecks in further increasing the apparent mobility and down-scaling the dimensions of the OFETs. To address this arduous challenge, Ohmic contacts at M/OSC interfaces with small (<100 Ω·cm) and voltage-independent $R_c$ are urgently desired.

The conductive channel in an OFET is confined in the first few molecular layers (MLs)

near the dielectric-semiconductor interface.[6] In the staggered structure devices, the contact resistance is the combination of two major components: (i) the resistance close to the M/OSC interface (referred to as $R_{int}$), and (ii) the resistance across the OSC layer (namely the access resistance $R_a$).[7] Unlike amorphous thin films, crystalline organic semiconductor with regular lattice structure and molecularly flat surfaces are excellent candidates to develop well-defined interfaces with the metal electrodes.[8,9] Several approaches have been studied to improve the $R_{int}$ in OFETs, including tuning the work function of metal electrodes by using oxidation,[10] forming self-assembled monolayers treatments or inserting thin layers with large dipole moments [11-14], and employing thin insulator layers for Fermi-level depinning. [15,16] On the other hand, the tunability of $R_a$ is generally weaker as it would rapidly increase with the number of MLs in the semiconductor.[11] Due to the potentially large density of trap states and defects, it is generally believed that OFETs with a monolayer channel would have lower performance than devices with thicker channels ranging in thickness from a few to tens of MLs.[13] However, if the defects or traps in the organic crystals are well controlled during the deposition and become negligible, in principle, organic single crystals with monolayer (1L) thickness and excellent in-plane crystallinity should be the ultimate solution for the OFETs with low resistance Ohmic contacts.

In this work, solution-processed 1L-crystals of 2,9-didecyldinaphtho(2,3-b:2′,3′-f)thieno(3,2-b)thiophene ($C_{10}$-DNTT) with negligible defects are employed as the active layers in OFETs. By fabricating the OFETs in one single-crystalline domain, we eliminated the grain boundary effect in our devices. The non-destructive deposition of the metal contacts and molecularly flat interface between the Au electrodes and the 1L-crystals leads to Ohmic-contact properties with $R_c$ as small as 40 Ω·cm, while the thermally evaporated Au electrodes show orders of magnitude higher $R_c$. For the 1L-devices with transferred electrodes, the $R_c$ shows no dependency on the drain-source voltage and the temperature, which distinguishes the Ohmic contacts in our 1L-OFETs from Schottky contacts. The alkyl side chains of the OSC molecules establish thin tunneling barriers which facilitate the depinning of the Fermi level at the M/OSC interfaces and the direct tunneling of the carriers. The superior contact properties allow the 1L-OFETs to operate at $V_{DS}$ down to -0.1 mV without affecting the effective carrier mobility. With the intrinsic mobility of 12.5 cm$^2$V$^{-1}$s$^{-1}$ and small $R_c$, high-field and high-current operation of OFETs was further studied. A high on-current density of 4.2 µA/µm is achieved by the 1L-OFETs, with significant velocity saturation and channel self-heating effects.

**Results**

### Contact resistance of 1L and 2L-OFETs

Large-area solution-deposition of 1L and 2L (dual-layer) C$_{10}$-DNTT crystals was performed by a solution shearing method similar to our previous publications [17,18]. With precise control of the temperature (from 60 to 65°C) and the shearing speed (from 2 to 3 µm/s), 1L and 2L crystals with single-crystalline domains extending over several millimeters can be achieved. The transferred-electrode OFETs utilized in the current work have a short channel length down to 2 µm. All devices studied in this paper, unless specifically mentioned, are fabricated so that the lateral charge transport occurs along the a-axis (the high-mobility axis) of the crystals. The detailed methodology in determining the a-axis of the crystals by cross-polarized optical microscopy (CPOM) is described in Fig. S1. The Au source and drain electrodes were pre-deposited onto an OTS-treated SiO$_2$ surface and mechanically transferred from there onto the organic-semiconductor surface. Transistors with channel lengths ranging from 8 µm to 53 µm were defined by the transferred Au electrodes on the single-crystalline domains of 1L (**Fig. 1A and B**) and 2L (Fig. 1F and G) crystals to facilitate transmission line method (TLM) measurements in order to evaluate the contact resistance. Prior to the electrical measurements, these regions of the OSC were isolated from the excess regions to avoid errors in the TLM fitting caused by the drain-source fringe current and the gate-source leakage current.[19] Comparisons on transfer properties were made on the OFETs with shortest channel length for the 1L (8 μm channel length) and 2L-devices (9 μm channel length). Transfer curves in the linear regime of operation were measured at $V_{DS}$ = -1 V and showed subthreshold swings of 370 and 366 mV/dec for 1L and 2L-devices, respectively. The on/off ratios are in the order of $10^6$, and the hysteresis is 0.47 V and 0.51 V (Fig. 1C and H), respectively. The governing equation for the TLM method is shown in Equation 1.

$$R_{tot}W = R_c W + \frac{L}{\mu_0 C(V_G - V_{TH})} = W(R_c + R_{ch}) \quad (1)$$

The width-normalized total resistance ($R_{total} \cdot W$) as a function of the channel length calculated at particular overdrive voltages ($V_G$-$V_{TH}$, where $V_{TH}$ is the threshold voltage) for the 1L- and 2L-OFETs are shown in Fig. 1D and I, respectively. The contact resistance ($R_c \cdot W$) and the intrinsic mobility ($\mu_0$) are extracted from the linear fits and plotted as a function of the overdrive voltage in Fig. 1E and 1J. The $R^2$ values for all the linear fits are higher than 0.99, and the error bars in Fig. 1E and 1J correspond to the standard error. For the 1L-devices, the $R_c \cdot W$ is 275 Ω·cm at an overdrive voltage ($V_G - V_{TH}$) of -10 V, 75 Ω·cm at $V_G - V_{TH}$ = -80 V and 40 Ω·cm at $V_G - V_{TH}$ = -125 V, which is the lowest reported value for staggered OFETs. On the other hand, the $\mu_0$ shows no strong dependence on the gate bias and is essentially constant at 12.5 cm$^2$/V$^{-1}$s$^{-1}$ for gate-source voltages greater than the threshold voltage. For the 2L-devices, the $R_c \cdot W$ at $V_G - V_{TH}$ = -80 V is 186 Ω·cm, i.e., about 2.5-times larger than in the 1L-devices. This implies that the second ML induces an additional access resistance ($R_a$) compared with the 1L-devices. However, the fact that the intrinsic mobility of the 2L-devices (12.4 cm$^2$/V$^{-1}$s$^{-1}$) is similar to that of the 1L-devices shows that the carrier transport in the monolayer C$_{10}$-DNTT is not affected by trap states or defects within the single-crystal domain. This observation confirms the advantages of using high-quality monolayer OSC crystals in staggered OFETs.

The major benefit of using transferred Au electrodes is to minimize the damage to the ultrathin semiconductor crystals during the metal-deposition process. The uniqueness of the metal transfer method applied on the solution-processed 1L-$C_{10}$-DNTT can be seen in **Fig. 2** by comparing the apparent mobilities of monolayer OFETs and multilayer OFETs (ML number ≥ 2) fabricated using the same semiconductor.[20-29] The apparent mobility values of the our devices were calculated from the 1L and 2L-devices with 53-μm channel length. As the thermal damage in the active layer is eliminated, the monolayer and multilayer (2L case in our study) OFETs with the transferred electrodes show almost identical mobilities. Results of a TLM study on 1L-OFETs with thermally evaporated Au electrodes are summarized in Section S1 and Fig. S2. The damage on the organic-semiconductor crystals by the thermally evaporated metal induces a large $R_c$ that renders the TLM invalid. Besides eliminating the thermal damages, a high $\mu_0$ is also essential in suppressing the $R_c$ of OFETs; this can be confirmed by the negative correlation between $\mu_0$ and $R_c$ seen in Fig. S3.[19,30-34] The outstanding $\mu_0$ and $R_c$ values suggest our solution-sheared 1L-crystals and transferred electrodes can be a promising technique for the fabrication of high-performance OFETs.

### $R_c$ dependency on $V_{DS}$ and temperature

Thermionic emission and field emission are the two major mechanisms for charge injection across a metal-semiconductor interface, which lead to Schottky contacts and Ohmic contacts, respectively.[35] In the scenario of thermionic emission, increasing $V_{DS}$ can enhance the band-bending and lower the contact resistance. As a result, the saturation mobility (i.e., the apparent mobility at high $V_{DS}$) is typically larger than the linear mobility (i.e., the apparent mobility at low $V_{DS}$) for OFETs with such contacts.[1,4] Similarly, when the temperature increases, the thermionic emission over the barrier would also be increased and thus lead to a negative correlation between the $R_c$ and $T$. The thermal activation can be extracted from a temperature-dependent measurement in the case of thermionic emission. In the present 1L-devices, the TLM measurements were conducted from $V_{DS}$ = -1 V to -1 mV and temperature from 340 K to 100 K. The $V_{DS}$-dependent TLM results were summarized in **Fig. 3A**, with the detailed data and fittings shown in Fig. 1C to E and Fig. S4 to S9. The calculated $R_c$ values at full turn-on regime ($V_G - V_{TH}$ = -60 V) and subthreshold regime ($V_G - V_{TH}$ = -10 V) were plotted in Fig. 3B against different $V_{DS}$. It was found that the contact resistance has little dependency on applied $V_{DS}$ both in the full turn-on regime and the subthreshold regime. The temperature-dependent TLM results from $T$ = 340 K to 100 K were summarized in Fig. 3C, with the detailed data and fittings shown in Fig. S10 to S16. As the temperature-dependent tests were conducted in a vacuum cryogenic probestation instead in nitrogen environment, a thin layer of Cytop (~200 nm thick) was employed to encapsulate the complete 1L-devices. From the summarized $R_c$ against $1000/T$ plot in Fig. 3D, a thermally activated charge injection barrier (7.8 meV from an Arrhenius fitting) was found at the subthreshold regime. However, no obvious dependency of the $R_c$ on temperature at the full turn-on regime. It suggests the charge injection is mainly through field emission instead of thermionic emission when the channel is fully turned on. The packed alkyl chains of the $C_{10}$-DNTT crystals (thickness of 1.2 nm), as illustrated in Fig. S17, behave as a tunneling barrier layer and reduce the pinning effect.[16] With the carriers injected and extracted mostly by field emission, the Ohmic contacts of the 1L-devices can be verified.

**Low drain-source operating power of 1L-OFET**

Low-resistance Ohmic contacts can lead to low drain-source operating voltage, which is critical for portable and wearable electronics. As low-dielectric-constant and relatively thick (300 nm) $SiO_2$ were employed for the current study, we thus only focus on the power dissipation between the drain-source electrodes while operating the 1L-OFETs. **Fig. 4A** shows that the transistors exhibiting proper transfer characteristics for drain-source voltages from -1 V to -0.1 mV. The on/off ratio of the devices is maintained larger than $10^3$, even at a small $V_{DS}$ of -0.1 mV. In Fig. 4B, the apparent linear mobility extracted using Eq. 2 is plotted as a function of the drain-source voltage, confirming that the apparent mobility of the 1L-OFET is independent of the drain-source voltage. The output behavior of the 1L-OFET is linear and symmetric as shown in Fig. 4C. It implies the device can be applied and modeled as a highly linear resistance in driving and sensing circuits, where small voltage fluctuations in $V_{DS}$ could lead to corresponding $I_D$ changes. If the required current level is fixed, the high mobility and low contact resistance in an OFET can minimize the corresponding power consumption. For example, the drain-source power ($P = I_D V_{DS}$) can be as small as 0.1 pW to induce a readable current of 1 nA. Such property is desired in low-power small-signal sensing, for example in bio-electronics applications. As a reference, the electrical signals in human bodies such as ECG (Electrocardiography), EMG (Electromyography), and action potentials of neurons have $V_{p-p}$ (peak to peak voltage) of 0.5 mV to 100 mV and $f$ from DC to 10 kHz.[36] The low-power AC behavior of the 1L-OFET was evaluated by applying sine waves generated using a function generator between the source and drain contacts, with $V_{p-p}$ ranging from 10 mV to 1 V and $f$ from 1 Hz to 1kHz (Fig. 4D and E). The small AC signals successfully induce the corresponding AC current in the device.

$$I_{DS} = \frac{W}{L} \mu_{app} C_i \left( V_{GS} - V_{TH} - \frac{V_{DS}}{2} \right) V_{DS} \quad (2)$$

**Velocity saturation and channel self-heating effects**

In addition to the field-effect mobility, the channel-width-normalized on-state drain current

($I_D/W$) is a more compact parameter describing the capability of a transistor in transporting carriers and driving other electronic components such as OLEDs and sensors. A high $I_D/W$ can be achieved by combining a small channel length, a low contact resistance and a high intrinsic mobility. With the Ohmic $R_c$ and the high intrinsic mobility observed in the 1L-OFETs, a high $I_D/W$ can potentially be achieved by reducing the channel length. Three OFETs with channel lengths of 140 μm, 8 μm and 2 μm were fabricated using Au electrodes transferred onto the same monolayer single crystal (**Fig. 5A**). The $I_D/W$ of the three devices measured at $V_{DS}$ = -1 V and $V_G$ = -80 V is 0.054, 0.59, and 0.88 μA/μm, respectively (Fig. 5B). Compared with the device with a channel length of 140 μm, the devices with channel lengths of 8 μm and 2 μm show 10-times and 16-times larger $I_D/W$. The reason why $I_D/W$ does not increase proportionally with the inverse of the channel length is that the contribution of the contact resistance to the total resistance increases with decreasing channel length. When we applied larger $V_{DS}$ in order to further increase $I_D/W$, an interesting phenomenon was observed. The resulting $I_D/W$ at $V_G$ = -80 V of the three devices becomes 1.9, 4.2, and 3.9 μA/μm, respectively (Fig. 5C), even though the short channel effect made an earlier turn-on and thus larger $I_D$ at the subthreshold regime for the $L$ = 2 μm device. This means that for the device with a channel length of 140 μm, an increase of the drain-source voltage by a factor of 80 leads to an increase in $I_D/W$ by a factor of 35, for $L$ = 8 μm, increasing $V_{DS}$ by a factor of 60 increases $I_D/W$ by a factor of 7, and for $L$ = 2 μm, increasing $V_{DS}$ by a factor of 40 increases $I_D/W$ by a factor of only 4.5. Reducing the channel length from 8 μm to 2 μm results in an even smaller width-normalized drain current. The apparent linear mobility and the threshold voltage extracted from Fig. 5B (summarized in Fig. 5D) are applied in Eq. 2 to calculate the output curves of the three devices (shown as the dotted curves in Fig. 5E, F, and G). In principle, the simulated curves should fit well with the measured output curves, as the contact-resistance contribution has been taken into account in the values of the apparent mobility. For the device with a channel length of 140 μm, the simulated output curves are indeed in good agreement with the measured curves. However, in the case of the devices with channel lengths of 8 μm and 2 μm, the measured drain current at high override voltage ($V_G$ = -80 V) is substantially smaller than simulated drain current. This implies that the drain current of these transistors is limited by mechanisms other than the contact resistance. The $R_{total} \cdot W$ of the devices with channel lengths of 140 μm, 8 μm and 2 μm are 4300, 1400 and 1000 Ohm·cm from Fig. 5C, which means the $R_c$ (at $V_G - V_{TH}$ = -80 V) of 75 Ohm·cm has minor contributions of 1.8%, 5.2%, and 7.3%, respectively. We then performed atomic force microscopy (AFM) on the channel region of a device with a channel length of 8 μm before and after operating the devices with large drain currents (Fig. S18) to confirm that the 1L-crystals were not damaged by the high current.

The large discrepancy between the measured and the simulated drain currents occurs primarily when $V_{DS}$ is large and the channel length is small, i.e., for large lateral electric fields ($F$). For $V_{DS}$ = -1 V, we calculated the lateral electric fields of 0.007, 0.12 and 0.5 V/μm for the devices with channel lengths of 140 μm, 8 μm and 2 μm, respectively, and for $V_{DS}$ = -80 V, we calculated $E$ = 0.57 V/μm for the device with a channel length of 140 μm; for these small electric fields the agreement between the measured and the simulated drain currents is good. However, for the 8-μm device at $V_{DS}$ = -60 V ($F$ = 7.5 V/μm), the measured $I_D/W$ deviates from the calculation (Fig. 5F). And the 2-μm device at $V_{DS}$ = -40 V ($F$ = 20.0 V/μm) shows an even larger deviation (Fig. 5G). It is understood that the velocity saturation effect of FETs will occur at high lateral field and high current density. The drift velocity of the carriers stops to increase at higher lateral field and becomes saturated, so that the actual $I_D/W$ and apparent mobility decrease from the theoretical value. Based on Eq. 3, the drift velocity ($v_d$) of the device with channel length of 8 μm can be calculated from the $I_D$ output curve at $V_G$ = -80V (Fig. S19). The $Q_{eff}$ represents the averaged density of the accumulated charges and can be calculated by Eq. 4. The $V_{TH}$ value is extrapolated from the linear transfer curve (Fig. S19). Finally, the $v_d$ is plotted against the applied lateral field (excluding the potential drop at the contacts) in Fig. 5H. To test whether carrier-velocity saturation indeed occurs in our devices, the Coughey-Thomas (Eq. 5) model, which is widely adopted to described the velocity saturation effect, is applied to fit the data.[37,38]

$$v_d = \frac{I_D}{W Q_{eff}} \quad (3)$$

$$Q_{eff} = C_i \left(V_G - V_{TH} - \frac{V_{DS}}{2}\right) \quad (4)$$

$$v_d = \frac{\mu_{LF} F}{\left[1+\left(\frac{\mu_{LF} F}{v_{sat}}\right)^{\gamma}\right]^{1/\gamma}} \quad (5)$$

The $C_i$ (areal capacitance) and $W$ (channel width) are 11.5 nF/cm² and 200 μm. A $v_{sat}$ value of 6.08 ± 0.03 × 10⁴ cm/s at 340 K was acquired from the fitting (Fig. 5H). The velocity value is ~2 orders smaller than those for inorganic semiconductor materials (e.g. 0.7-1.0 × 10⁷ cm/s for Silicon, 3.4 × 10⁶ cm/s for 1L-MoS₂).[37,38] The fitted low-field mobility ($\mu_{LF}$) is 13.5 ± 0.54 cm²V⁻¹s⁻¹, which agrees well with the intrinsic mobility value measured at $V_{DS}$ = -1 V (Fig. 1G). The fitted $\gamma$ at 340 K is 1.49 ± 0.06, and, as a comparison, the $\gamma$ of Si is ~2 for electrons and ~1 for holes. For the same device test at 100 K, an increase in the $v_{sat}$ is observed (9.26 ± 0.15 × 10⁴ cm/s). Given the more negative $V_{TH}$ (-21 V and -11.5 V at 100 K and 340K respectively in Fig. S19A), the $Q_{eff}$ at 100 K should be lower than at 340 K. However, the output current at 100 K (Fig. S19D) is even larger than at 340 K (Fig. S19C). It suggests lower temperature is beneficial for higher $v_{sat}$ and thus higher current, which is in agreement with the studies on Si and MoS₂ semiconductor.[37,38] Since the $v_{sat}$ of an OFET device is related to the temperature, the high-current operation of an OFET may induce a channel self-heating effect and in-turn limit the current from increasing any more. It is thus critical to know the temperature of the channel during operation. Take the 8-μm device for example, as the channel dimension is 200 × 8 μm², it is difficult for typical thermocouple measurement or infrared mapping. The finite element model for heat transfer is thus utilized to back out the temperature during operation

(**Fig. 6A**). The detailed modeling parameters can be found in Section S2. For the 8-μm device operated at $V_{DS}$ = -60 V and $I_D/W$ = 4.2 μA/μm (corresponding to $P_{channel}$ = $V_{DS}I_D/WL$ = 31.3 W/mm$^2$), the $\Delta T$ of the channel is up to 14.53 K (Fig. 6B and C). On the contrary, the $\Delta T$ is only 0.034 K at $V_{DS}$ = -1V (Fig. S20), which can explain why the linear operation of the 8-μm device agrees well with the standard model, but the saturation operation does not. As a comparison, we employed a micro-thermocouple to record the real-time temperature of a 1L-OFET operated at high current densities (Fig. S21). Despite the large thermal mass effect of the thermocouple, obvious $\Delta T$ is observed at high $V_{DS}$ bias (-40 V and -10 V). On the other hand, the low $V_{DS}$ (-1 V) bias does not induce readable $\Delta T$ to the thermocouple.

As shown in Eq. 3, when velocity saturation occurs, the $I_D$ of an FET is limited by $v_{sat}$ instead of $\mu_0$ or $L$. In fact, further shortening the $L$ may lead to a larger $P_{channel}$ and thus a more significant self-heating effect. Since the $v_{sat}$ is found negatively correlated with $T$, shortening the $L$ may possibility decrease the $I_D$ rather than increase it. The simulated $\Delta T$ of the 2-μm device is 17.75 K (at $V_{DS}$ = -40 V), higher than the case of 8-μm device. To a certain extent, the self-heating effects limits the achievable $I_D/W$ of OFETs. It is noteworthy the $\Delta T$ is simulated based on single device. The heat accumulation will become more severe when we try to down-scale the OFETs and put more devices together. As the channel length is decreased to 100 nm and maintained the same power density ($P_{channel}$ = 31.3 W/mm$^2$), the heat accumulation of the 250 devices makes the temperature increase for 86 K, i.e. 5.9 times of the single device case (Fig. 6D and E). It suggests there will be a great challenge when large number of high-performance short-channel devices are integrated together. More importantly, the heating effect is evaluated on SiO$_2$/Si substrate with a high thermal conductivity ($k_{SiO2}$ = 1.4 W/mK, $k_{Si}$ = 130 W/mK), i.e. good heat dissipation property. By replacing the SiO$_2$/Si substrate with 5-μm-thick poly(ethylene paraphthalate) (PET, $k_{PET}$ = 0.4 W/mK), the $\Delta T$ is over 6 times more significant compared with the SiO$_2$/Si case (Fig. 6F). Even with high-mobility crystals and ohmic contact, the channel self-heating effects will limit the current to only 1/6 of the value on SiO$_2$/Si substrate, just because the heat generated during operation cannot be dissipated. It is thus of great importance to study the electro-thermal properties and behaviors of OFETs before the realization of practical high-performance flexible electronics.

**Conclusion**

In conclusion, we have shown that the monolayer single crystals are the optimum form of active layers for the staggered OFETs in terms of high mobility and low contact resistance. The generally reported lower mobility of monolayer OSCs compared with their thicker counterparts in the literature may arise from the thermal damage caused by the electrode deposition process. $R_c \cdot W$ as low as 40 Ω·cm and $\mu_0$ of 12.5 cm$^2$V$^{-1}$s$^{-1}$ are achieved by the 1L-OFETs. The non-destructive Au/OSC interfaces and the aligned alkyl chains effectively facilitate the depinning of the Fermi level. The charge injection is mainly through field emission instead of thermionic emission once the channel is turned on, proving an apparent Ohmic contact of the 1L-devices. The 1L-OFET operates linearly from $V_{DS}$ = -1 V to $V_{DS}$ as small as -0.1 mV. The low contact resistance and high mobility allows the 1L crystal to transport current density as high as 4.2 µA/µm. At such high current density, the velocity saturation and channel self-heating effects are first studied for planar OFETs, indicating the superior 1L-OFETs are approaching the physical limitations of organic semiconductor materials under room temperature and DC operation. With the high mobility (both intrinsic and extrinsic) and low contact resistance, the 1L-crystal based OFETs are critical corner stones for the OFETs to achieve high-density, high-speed flexible electronics in the coming future. But the thermal-related properties of the OFETs during operation should be the next major challenge in front of that goal.

**Methods:**

*Solution-shearing of 1L-crystal*

Si wafers (525 μm thick, Namkang Hi-Tech) with 300-nm-thick thermal oxide layer were cleaned by oxygen-plasma (30 W, Harrick Plasma) for 30 min. The wafers were loaded into a vacuum oven with 50 μL PTS (J&K Scientific) aside. The oven was heated to 150°C, stayed at 150°C for 60 minutes, and cooled to room temperature. The RMS roughness of the PTS-treat $SiO_2$/Si wafers was 0.28 nm (5 μm × 5 μm area). During the solution-shearing process, the PTS-treated substrate and OTS (Sigma Aldrich) -treated blade (also $SiO_2$/Si) were heated up to 60°C. 40 μL $C_{10}$-DNTT solution (0.2 mg/ml in tetralin, heated to 70°C to help dissolving) was inject between the substrate and blade. The shearing rate was controlled by a linear translation stage (ILC 100 CC, New port) at 2.5 μm/s. After the deposition, the samples were stored in a vacuum oven (OV-12, Jeio Tech) for at least overnight to remove residual solvent. Before the electrode deposition or transfer, the 1L-crystals were transferred into a glovebox (water and oxygen content lower than 1 ppm, MBraun) and heated to 80°C for 15 min.

*Transfer of Au electrodes*

The Au electrodes (180 nm thick) were formed by thermal evaporation on OTS-treated $SiO_2$/Si wafers. The resulting Au stripes had a length of 200 μm and width of 35 μm. The Au stripes were transferred by a probe station equipped with a microscope. When the stripes were to attach the surface of 1L-crystal, the length of the stirpes were controlled to be perpendicular to the a-axis of the crystal. All the electrodes for OFETs with different channel length in the TLM configuration were ensured to be in a single crystal. After the attachment of Au electrodes, the testing area of 1L-crystal were carefully separated with the excess area by a rigid probe to minimize the fringe current. Then the completed devices were transferred to the glove box, heated to 80°C for 15 mins, and stored at room temperature for overnight.

*Electrical characterization of the OFETs*

All measurements were performed in the glove box environment and in the dark. A dual-channel sourcemeter (Keithley 2636A) was employed to test the transfer and output characters of the OFETs. The voltage scanning rate was 10V/s for the forward and reverse scan. The AC voltage signals were supplied by a function generator (33210A Keysight), and the AC current were measured by a high-speed current meter (DMM 6500 Keithley). To ensure proper electrical contacts between the probes and the transferred electrodes. Au wires (15 μm diameter) were attached on the ends of the probes. The TLM was performed by linear fitting of total channel resistance (width normalized) of OFETs with difference channel length at the same $V_{DS}$ and ($V_G - V_{TH}$). The error bars in the TLM plots represent the resistance different in the forward and reverse bias (i. e. hysteresis). The error bars in the $R_c$ plots represent the total absolute errors of the TLM fitting.

**Acknowledgements**

We gratefully acknowledge the support from General Research Fund (GRF) under Grant No. HKU 17264016 and HKU 17204517. We also thank Dr. Hagen Klauk and James W. Borchert for the fruitful discussions and suggestions.

**Author contributions**

B. P. designed and performed the experiments and analyzed the data. H. Y. L contributes on the simulation of channel self-heating. M. C. helps in the crystal fabrication. B. P. and P. K. L. C. wrote the manuscript. P. K. L. C supervised this work. All authors discussed the results and reviewed the manuscript.

**Additional information**

**Supplementary Information** accompanies this paper at
http://www.nature.com/naturecommunications

**Competing financial interest:** The authors declare no competing financial interests.

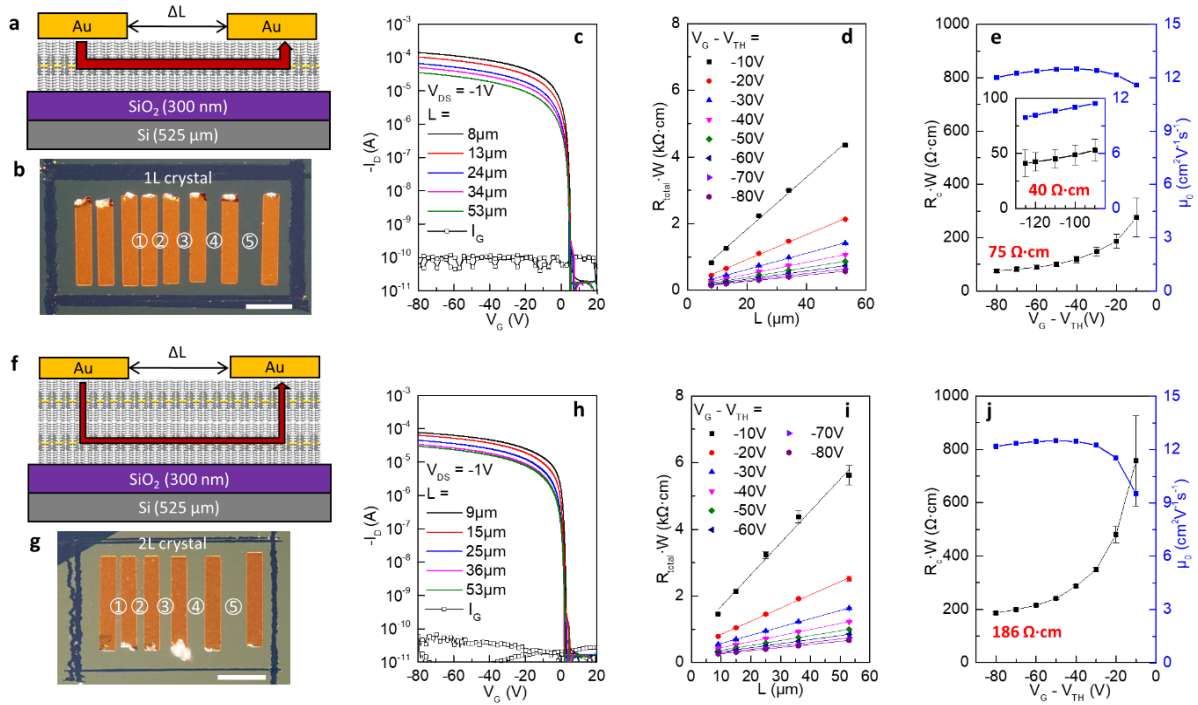

**Fig. 1 | Contact resistance of field-effect transistors based on 1L and 2L crystals of the small-molecule organic semiconductor $C_{10}$-DNTT.** Schematic cross-sections of bottom-gate, top-contact (BGTC) OFETs based on (**a**) a 1L-crystal and (**f**) a 2L-crystal of $C_{10}$-DNTT as the active semiconductor layer. Cross-polarized optical-microscopy (CPOM) images of OFETs fabricated by transferring Au source and drain electrodes (electrode dimensions: 200 μm × 35 μm) onto (**b**) a 1L-crystal and (**g**) a 2L-crystal of the organic semiconductor. The active regions of the transistors have been isolated by scratching using a sharp tungsten probe needle. The scale bars in (**b**) and (**g**) represent 100 μm. Transfer curves measured in the linear regime of operation and plotted on a semi-logarithmic scale of (**c**) the 1L-devices and (**h**) the 2L-devices. Total device resistance calculated using the transmission line method (TLM) and plotted as a function of the channel length of the transistors for various gate overdrive voltages ($V_G - V_{TH}$) for (**d**) the 1L-devices and (**i**) the 2L-devices. Calculated contact resistance (black squares) and intrinsic mobility (blue squares) at various gate overdrive voltages ($V_G - V_{TH}$) for (**e**) the 1L-devices and (**j**) the 2L-devices.

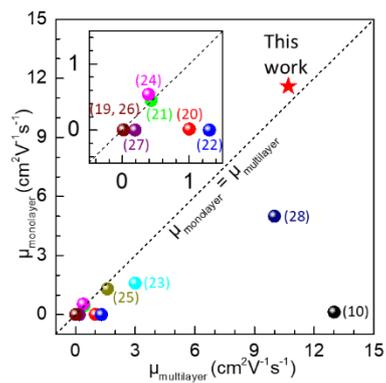

**Fig. 2 | The performance of 1L-OFET compared with literatures.** A summary of the reported apparent mobility based on the monolayer and multilayer of the same OSC materials. Black dashed line represents the monolayer and multilayer OSC having identical mobility.

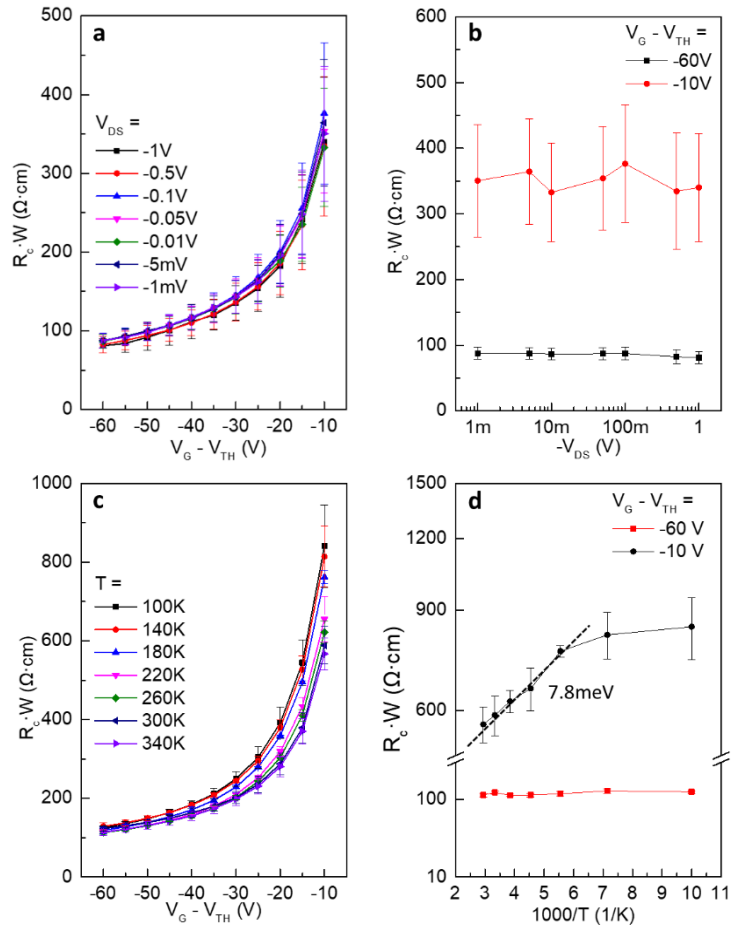

**Fig. 3 | The contact resistance at different $V_{DS}$ and temperature.** (**a**) The TLM contact resistance of 1L-OFETs at different $V_{DS}$ from -1 V to -1 mV. (**b**) The contact resistance against $V_{DS}$ at the full turn-on regime ($V_G$ - $V_{TH}$ = -60 V) and the subthreshold regime ($V_G$ - $V_{TH}$ = -10 V). (**c**) The TLM contact resistance of 1L-OFETs at $V_{DS}$ = -1 V and temperature from 100 to 340 K. (**d**) The contact resistance against 1000/$T$ at the full turn-on regime and the subthreshold regime. The black dashed line indicates an Arrhenius fitting of 7.8 meV.

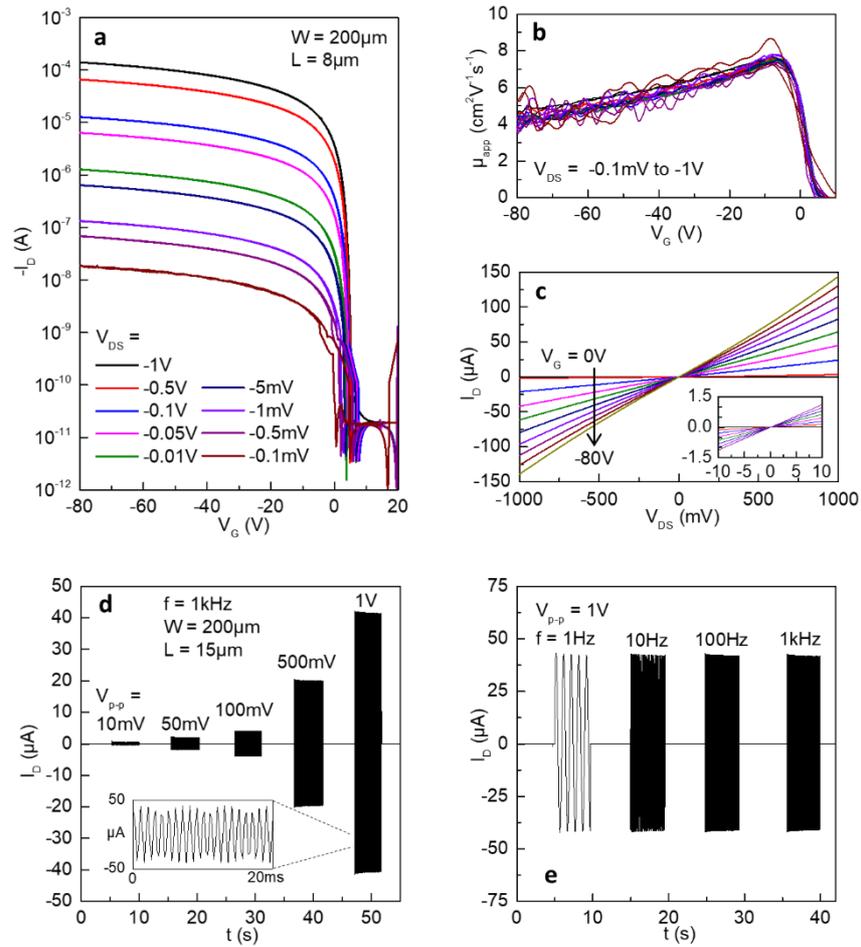

**Fig. 4 | Low drain-source operating power of the 1L-OFETs.** (**a**) The linear transfer curves of an 1L-OFET with 8-μm channel length at different $V_{DS}$ from -1V to -0.1 mV. (**b**) The apparent linear mobility of the device in (**a**). (**c**) The output curves scanning from $V_{DS}$ = 1 V to -1 V of the device in (**a**). Inset is the output scanning from $V_{DS}$ = 10 mV to -10 mV. (**d**) AC $I_D$ responses at AC $V_{DS}$ input of sine waves with frequency of 1 kHz and $V_{p\text{-}p}$ (peak to peak voltage) from 10 mV to 1 V. Inset is the zoom-in plot at $f$ = 1 kHz and $V_{p\text{-}p}$ = 1 V. (**e**) AC $I_{DS}$ responses at AC $V_{DS}$ input of sine waves with $V_{p\text{-}p}$ of 1 V and $f$ from 1 Hz to 1 kHz.

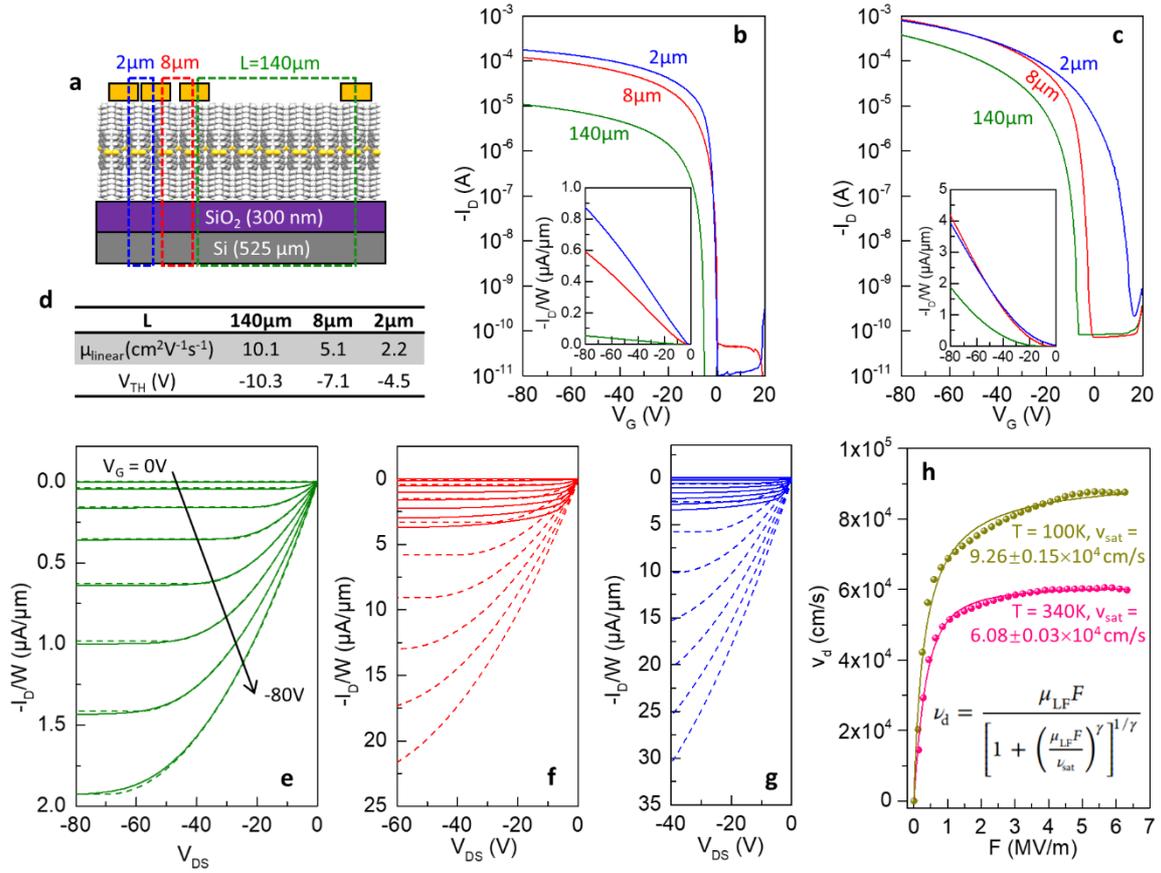

**Fig. 5 | The velocity saturation effect of short-channel 1L-OFETs.** (**a**) The schematic illustration the 2-μm, 8-μm, and 140-μm devices on a 1L single crystal. (**b**) The linear transfer curves ($V_{DS}$ = -1V) of the 140-μm (green), 8-μm (red), 2-μm (blue) devices. Inset is the $I_D/W$ calculated from $I_D$. (**c**) The transfer curves of the 1L-devices at large $V_{DS}$. $V_{DS}$ = -80 V for 140-μm device (green), -60 V for 8-μm device (red), and -40 V for 2-μm device (blue). Applying larger $V_{DS}$ bias may cause breakdown for the 8-μm and 2-μm devices. Inset is the $I_D/W$ calculated from $I_D$. (**d**) The summary of the $\mu_{linear}$ and $V_{TH}$ calculated from (**b**). The simulated output curves (dashed lines) and measured output curves (solid lines) of the (**e**) 140-μm, (**f**) 8-μm, (**g**) 2-μm devices at $V_G$ from 0 V to -80 V. (**h**) The fittings on drifting velocity (as calculated from the measured output curves) by Caughey-Thomas model for the 8-μm device measured at 340 K and 100 K. The drain output current at $V_G$ = -80 V is utilized for the fitting.

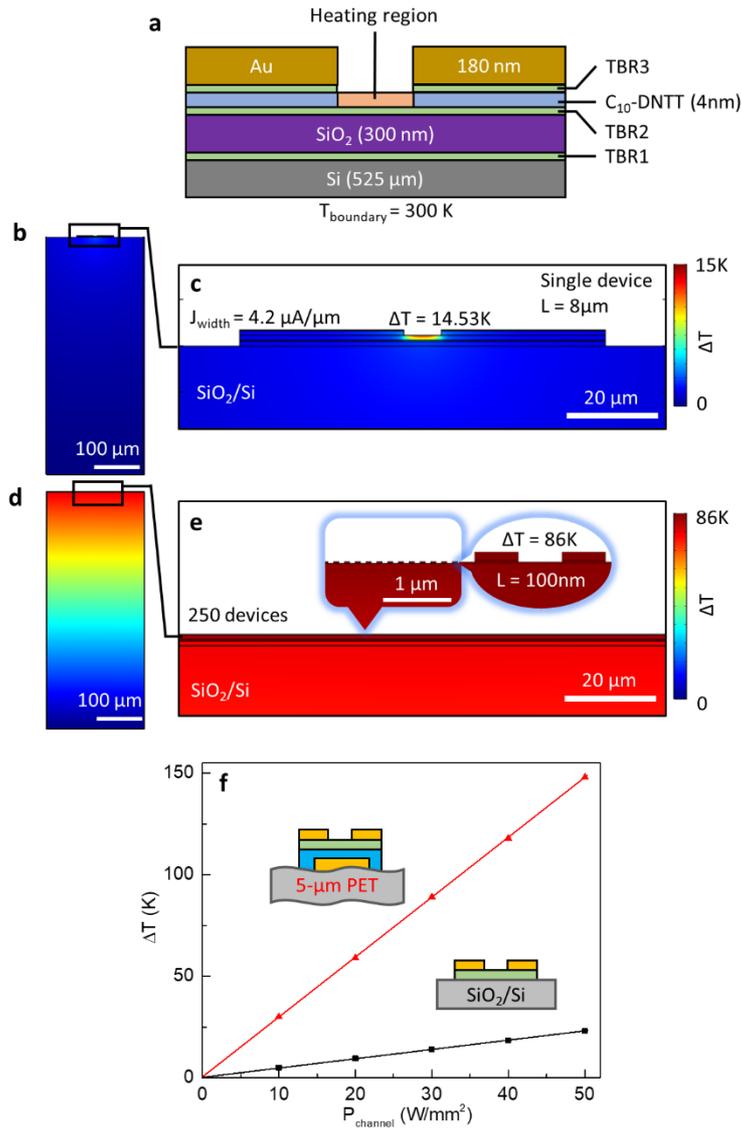

**Fig. 6 | Channel self-heating effect of 1L-OFET by finite-element simulation.** (**a**) The schematic layered structure of the simulated model. TBR (thermal boundary resistance) was modeled by a 1-μm-thick layer with fixed thermal resistance. The bottom boundary was fixed at 300 K. Other boundaries were set as air convection ($h = 10^4$ W/m$^2$K). (**b**) The finite simulation result and (**c**) the enlarged view of the temperature mapping for channel self-heating of the 8-μm device. (**d**) The heat accumulation effect and (**e**) the enlarged view of the temperature mapping from 250 devices with short channel length and high density. (**f**) The simulated maximum $\Delta T$ versus input power. The red line represents the 5-μm-thick PET film as the substrate. The black line is the SiO$_2$/Si as the substrate.